\documentclass[12pt,english]{article}
\pdfoutput=1
\usepackage[T1]{fontenc}
\usepackage{fullpage}
\usepackage{authblk}
\usepackage{graphicx,array}
\usepackage{cite}
\usepackage{color}
\usepackage{latexsym}
\usepackage{amsthm}
\usepackage{amsmath}
\usepackage[titletoc]{appendix}
\usepackage{enumitem}
\usepackage{amssymb}
\usepackage{color}
\usepackage{authblk}
\usepackage[unicode=true,
 bookmarks=true,bookmarksnumbered=false,bookmarksopen=false,
 breaklinks=false,pdfborder={0 0 1},backref=false,colorlinks=true]
 {hyperref}
\usepackage{listings}
\usepackage{mathrsfs}
\DeclareMathOperator{\Tr}{Tr}

\newcommand{\bra}[1]{\langle #1 |}
\newcommand{\ket}[1]{| #1 \rangle}

\newcommand{\be}{\begin{equation}}
\newcommand{\ee}{\end{equation}}
\theoremstyle{definition}
\theoremstyle{plain}
\newtheorem{thm}{Theorem}[section]

\newtheorem{prop}[thm]{Proposition}

\theoremstyle{definition}
\newtheorem{defn}{Definition}[section]

\usepackage{color}
\hypersetup{pdftitle={title}, pdfauthor={}, citecolor=blue,linkcolor=blue,urlcolor=blue,citecolor=blue}

\begin{document}
\vspace{2cm}
\thispagestyle{empty}
\begin{center}
{\LARGE \bf
Minimal Purifications, Wormhole Geometries, and the Complexity=Action Proposal
}\\
\bigskip\vspace{1cm}{
{\large Ning Bao,${}^a$}
} \\[7mm]
 {\it ${}^a$Center for Theoretical Physics and Department of Physics \\
     University of California, Berkeley, CA 94720, USA and \\
     Lawrence Berkeley National Laboratory, Berkeley, CA 94720, USA \\[1.5mm]} \let\thefootnote\relax\footnote{\noindent e-mail: \url{ningbao75@gmail.com}} \\
 \end{center}
\bigskip
\centerline{\large\bf Abstract}
\begin{quote} \small
We define a new criterion for selecting a specific minimal entanglement purification of given mixed states in generic quantum states using the entanglement of purification. We then propose that its holographic dual is the state living on the boundary of the entanglement wedge in the surface-state correspondence. Finally, we make some remarks about the relationship between this and the complexity equals action proposals.
\end{quote}

\section{Introduction}
The discovery of the Ryu-Takayanagi formula \cite{ryu2006holographic} giving the entanglement entropy of some boundary subregion $A$ has had profound implications for the study of quantum gravity and the AdS/CFT correspondence. The formula is given by
\begin{equation}
S(A)=\frac{Area}{4G},
\end{equation}

Where the area is that of the minimal surface in the bulk homologous to the boundary subregion. The existence of this relation has allowed for the proof of many constraints \cite{hayden2013holographic, bao2015holographic} that distinguish holographic states dual to semi-classical spacetime geometries from generic quantum states, and has offered insights into the precise nature of how spacetime can emerge from entanglement.

Entanglement entropy is not a sufficient measure of entanglement by itself, however. For example, while the distillable entanglement between two bipartitions of a pure state is given by the entanglement entropy of either of the two subsystems, the same cannot be said for bipartitions of a mixed state. To make further progress in the direction, the entanglement of purification, discovered by \cite{terhal2002entanglement}, was proposed as an entanglement measure which is calculated by purifications of a mixed state which retain the entanglement structure across the given bipartition of the mixed state. The holographic dual of this object has been conjectured in \cite{takayanagi2017holographic, nguyen2018entanglement}, and has been generalized to multipartite and conditional contexts in \cite{bao2018holographic, bao:2018gck, umemoto2018entanglement} and applied in an interesting fashion in \cite{nomura2018pulling}.

Further, we note that while it is well known that the purification of a given mixed state is unique up to ancilla padding and application of a unitary to the purifying subsystem, the picking out of a specific unitary, or, equivalently, a specific purification, is a process that is not yet fully understood.

In this work, we use the entanglement of purification as a constraint condition to identify uniquely a specific purification of a given mixed state which we will define to be the minimally entangled purification. We will prove the uniqueness and existence of this purification as defined, and we will show that it is dual to the state defined on the boundary of the entanglement wedge as defined by the surface-state correspondence \cite{Miyaji:2015yva}. We will conclude with some comments on how this can be related to the complexity=action proposal \cite{Brown:2015lvg}.

\section{Purifications of Mixed States}
Consider a mixed state $\rho_A$ in any Hilbert space dimension. Such a mixed state can always be embedded in a pure state $\ket{\Psi}_{AB}$ for some purifying system $B$ such that $\Tr_B \ket{\Psi}\bra{\Psi}_{AB}=\rho_A$. This embedding is, however, highly nonunique: first, one is free to add arbitrary numbers of unentangled pure ancilla systems to a given purifying $B$ subsystem, as doing so  maintains the status of the augmented system as a purifying system:

\begin{equation}
\ket{\Psi}_{AB} \rightarrow \ket{\Psi}_{AB}\otimes \ket{000...}.
\end{equation}

Without loss of generality we can take these ancilla systems to be product states of unentangled qubits, as will soon become apparent. With this restriction, we see that this method of generating purifications through unentangled ancilla qubits is unique up to the number of ancilla qubits used.

Second, one is free to act with arbitrary unitary matrices on the potentially ancilla-augmented $B$ subsystem:

\begin{equation}
\ket{\Psi}_{AB} \rightarrow U_B\ket{\Psi}_{AB}.
\end{equation}

So long as these unitary operations do not also act on the $A$ subsystem and the $B$ system purifies $\rho_A$, the new system will also purify $\rho_A$. This unitary freedom allows for even more freedom in the choice of purifications of $\rho_A$, even whilst keeping the Hilbert space dimension of the purification constant. It also allows for consideration of only unentangled qubits as the ancilla system, as other ancilla systems can be constructed by acting with unitaries on product states of unentangled qubits. Indeed, unitaries that mix the original $B$ system with the ancilla qubits are also allowed.

With such large freedomsin the choice of purification of a given density matrix, it is natural to ask whether there are purifications which can be distinguished from the rest. It is trivially true that there is no ``largest possible'' purification, as one can always introduce additional unentangled ancilla qubits via tensor product. There is, however, a smallest one: the purifying system of minimal Hilbert space dimension precisely equal to $e^{S(A)}$, where $S(A)$ is the entanglement entropy of $\rho_A$ given by

\begin{equation}
S(A)=-\Tr \rho_A \log \rho_A.
\end{equation}

This is made clear by considering entanglement distillation: the entanglement between $\rho_A$ and its purification can be distilled into a number of Bell pairs equal to $S(A)$ via LOCC asymptotically without ever mixing the two systems. \footnote{Here we make only asymptotic statements, assuming a large number of copies of $\rho_A$, as the one-shot version of these statements will allow for distillation of a number of qubits equal to the min-entropy.} Here, it is clear that the density matrix for the $B$ subsystem is a maximally mixed state with Hilbert space dimension $e^{S(A)}$. \footnote{Technically this should be the floor of $e^{S(A)}$ to ensure an integer number of e-bits.} As the purification is a maximally mixed state, it cannot be made any smaller in Hilbert space dimension without violating the fact that the entanglement entropy of a subsystem is upper bounded by the logarithm of its Hilbert space dimension. From here, one can simply apply the inverse unitary on the $A$ system to recover $\rho_A$, while keeping the $B$ system the same. This shows the claim above.

This argument, however, only fixes the freedom in the choice of the Hilbert space dimension of the purification. One is still free to apply unitaries to the $B$ subsystem of $\ket{\Psi}_{AB}$. Note that while this will not change $\rho_B$, it will in general change which portions of the $A$ system are purified by which portions of the $B$ system, assuming a way of differentiating the Hilbert space factors in the $B$ system. Consider, for example, the $AB$ state to be a pair of Bell pairs, and the unitary in question one that permutes the $B$ halves of the Bell pairs. In order to fix the minimal purification up to the choice of unitary, further constraints are required.

A relevant quantity to consider in constraining such purifications is the entanglement of purification, $E_P(A_1:A_2)$, first defined in \cite{terhal2002entanglement}. Further properties of this quantity were discussed in \cite{bagchi2015monogamy} Consider a factorization of the $A$ system into two subsystems, $A_1$ and $A_2$. The entanglement of purification between the $A_1$ and $A_2$ subsystems is given by

\begin{equation}
E_P(A_1:A_2)=\inf_{A'_1A'_2} S(A_1A'_1),
\end{equation}

Where $A'_1A'_2$ jointly purify $\rho_{A_1A_2}$, and there is some Hilbert space decomposition into an $A_1A'_1$ and $A_2A'_2$ Hilbert space subfactors. This is a different notion of minimization over possible choices of purifications, as this minimizes the entanglement between the subsystems of the $A$ system even in the augmented Hilbert space subfactors they are embedded in within the full pure state $\ket{\Psi}_{AB}$. Note that this breaks the degeneracy in the two Bell pair example before: if $A'_1$ is chosen to be the Bell pair half that purifies $A_1$, $S(A_1A'_1)$ would be zero, but if $A'_1$ is chosen to be the Bell pair half that purifies $A_2$, then $S(A_1A'_1)$ is clearly nonzero, even though these choices are related by unitaries of the same Hilbert space dimension. Note, however, that this analysis is insensitive to padding of the purifying system with unentangled ancilla qubits, as this would not change $S(A_1A'_1)$. Therefore, the first condition, the constraint on Hilbert space dimension of the purifying system given by $S(A)$, is independently required. 

Moreover, there is still a degeneracy: consider the case where $A_1$ is three Bell pair halves, and $A_2$ is another uncorrelated three Bell pair halves. The optimal purifying system here is clearly the six purifying Bell pair halves, separated into the three that purify $A_1$ and the three that purify $A_2$, respectively, to give zero $E_P(A_1:A_2)$. Note, however that unitaries that permute the three Bell pair halves corresponding to $A'_1$ or those of $A'_2$ do not change the entanglement of purification, though they would in general change a different entanglement of purificaiton corresponding to a division of $A$ into different subsystems.

This motivates a notion of a minimal purification: the purification that minimizes all entanglements of purifications of all bipartitions of $\rho_A$ simultaneously, with minimal Hilbert space dimension.

\begin{defn}Given a density matrix $\rho_A$, a minimal entanglement purification of $\rho_A$ is a purification $\ket{\Psi}_{AB}$ such that $\log \dim H_B=S(A)$ and that satisfies the infimum over all purifications in the definition for the entanglement of purification for all bipartitions of $\rho_A$ into $A_1$ and $A_2$ subsystems.\end{defn}

Such a definiton is very strongly constrained, and so we therefore propose its existence and uniqueness \footnote{We note that for this proposition and the following proof notions of asymptotic instead of one-shot distillation should be considered, with the appropriate epsilons and errors involved in such}.

\begin{prop}The minimal entanglement purification $\ket{\Psi}_{AB}$ for a given $\rho_A$ exists and is unique.\end{prop}

The proof of this proposition proceeds as follows. Consider, first, an arbitrary purification of $\rho_A$, $\ket{\Psi}_{AB}$; such a purification is guaranteed to exist. We will whittle down this purification to a more minimal one. Consider once again the distillation of the entanglement between $A$ and $B$ by LOCC operations, converting the entanglement between $A$ and $B$ into Bell pairs. This leaves three systems: $\bar{A}$, $\bar{B}$, and the Bell pairs between $A$ and $B$, where the barred systems are the parts of $A$ and $B$ that are unentangled with $B$ and $A$, respectively. By the minimal Hilbert space conditon on the purification the $\bar{B}$ system is trivial.

Now, in this context, let us consider every bipartition of $A$ into subsystems $A_1$ and $A_2$. Without loss of generality, let us consider only $A_1$. $A_1$ contains subsystems of both $\bar{A}$ and its Bell pair halves purified by $B$, called $A'_1$. Its $S(\bar{A})$ component is the irreducible entanglement of purification, i.e. the thing that would always lower bound $S(A_1A'_1)$ as the introduction of $B$ does not affect it. If we take the Bell pair contribution to $A_1$ to be purified by precisely what we define to be $A'_1$, \footnote{This is something that we are free to choose.} then the Bell pair contribution to the entanglement of purification is zero. Subadditivity of the entanglement entropy,
\begin{equation}
S(\bar{A}_1A'_1)\leq S(\bar{A}_1)+S(A'_1),
\end{equation}
Therefore giving that the entanglement of purification comes only from the $\bar{A}$ contribution to $A_1$, which is data given from $\rho_A$. As this argument also fixes the partitioning between $A'_1$ and $A'_2$ in the purification, this argument proves the existence of the minimal entanglement purification.

For uniqueness, one need simply note that the number of independent entanglement of purifications that need to be fixed map precisely onto the Bell pairs distillable between $A$ and $B$. Because of the requirement that $\ket{\Psi}_{AB}$ calculates all entanglements of purifications, this uniquely fixes every $B$-associated Bell pair half with an $A$-associated Bell pair half. Because the minimal purifying system is only the system of these Bell pair halves in $B$, this uniquely specifies the $B$ subsystem and its correlation structure with $A$, thus fixing $\ket{\Psi}_{AB}$ to be unique.

\section{Holographic Minimal Entanglement of Purification}
The previous notion is sharpened in the context of a conjecture regarding the holographic dual of the entanglement of purification called the $E_P=E_W$ conjecture \cite{takayanagi2017holographic, nguyen2018entanglement}. Consider a holographic conformal field theory, where $A$ is a subregion fo the boundary with a nontrivial entanglement wedge. That conjecture states that the holographic dual of the entanglement of purification is the are of the entanglement wedge cross section, defined to be the minimal surface partitioning the entanglement wedge of $\rho_A=\rho_{A_1A_2}$ into two regions, one adjacent to $A_1$ but not $A_2$ and one adjacent to $A_2$ but not $A_1$. 

\begin{figure}[h]
\caption{The entanglement wedge cross section of the $A_1A_2$ boundary subsystem, with size $E_W$.
This object, depicted as a red, dashed line, is the minimal surface that totally partitions the entanglement wedge into a region adjacent to $A_1$ and one adjacent to $A_2$. The $A'_1$ system is then taken to live on the portion of the RT surface on the $A_1$ side of the cross section, and the $A'_2$ is taken to live on the $A_2$ side.}
\centering
\includegraphics[width=0.5\textwidth]{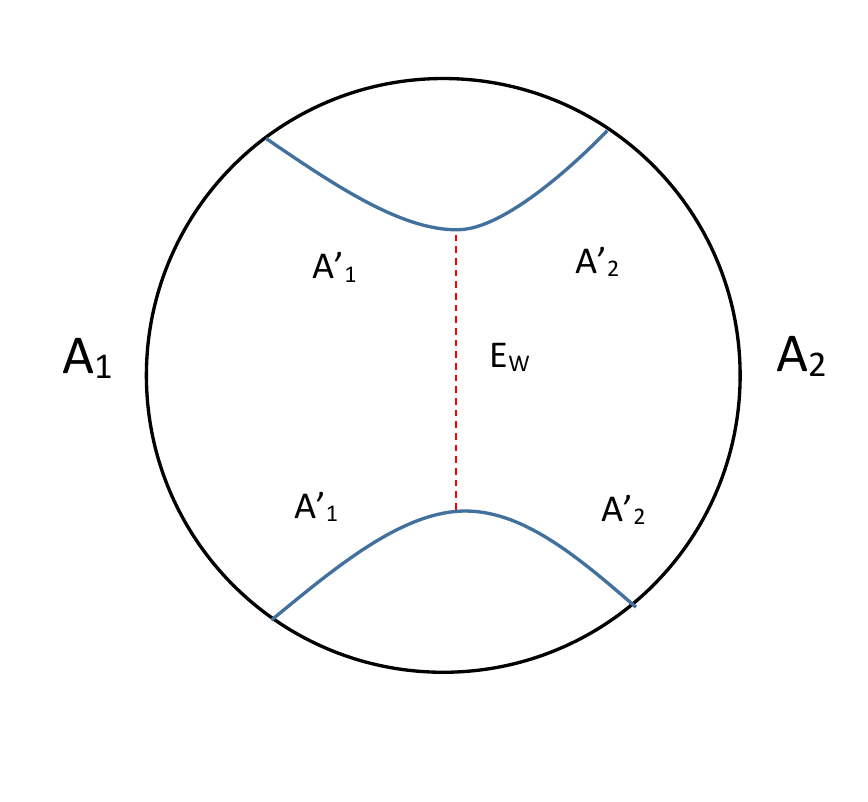}
\label{fig:2pep2}
\end{figure}

Further, it exploits the surface-state correspondence \cite{Miyaji:2015yva} \footnote{The surface state correspondence \cite{Miyaji:2015yva} is a conjectured correspondence motivated by tensor networks that states in part that one can use isometries from the boundary CFT to any homologous surface to define a state on that bulk homologous surface of the same purity as the boundary state. Here, in particular, we take the homologous surface to the entire boundary to be the boundary of the entanglement wedge of $\rho_A$.} to posit that the $A'_1$ and $A'_2$ Hilbert space subfactors live on the Ryu-Takayanagi (RT) surfaces bounding the entanglement wedge of $\rho_{A_1A_2}$, doing so by invoking an isometry mapping from the portion of the boundary complement to $A_1A_2$ to the RT surfaces such that the state defined on the union of $A_1A_2$ and its associated RT surface has the same purity as the state on the entire boundary. Let us restrict to the case where the boundary state is a pure state. Moreover, because the RT surfaces are precisely large enough to hold a purifying system of Hilbert space dimension equal to $e^{S(A)}$, as its area is directly proportional to the entanglement entropy, \footnote{This assumes a constant density of degrees of freedom defined on the RT surface motivated by the bottleneck arguments of \cite{freedman2017bit}.} the purification of $A_1A_2$ living on the RT surfaces is the one of minimal Hilbert space dimension.

In order for the $E_W$ surface to be the geometrization of the entanglement of purification, it must also be the bottleneck for the maximum number of bit threads as defined in \cite{freedman2017bit} emerging from the $A_1A'_1$ subsystem to the $A_2A'_2$ subsystem constrained to go only through the entanglement wedge. \footnote{Bit threads \cite{freedman2017bit} are a flow reformulation of the Ryu-Takayanagi formula. If one assumes the surface-state conjecture, then it naturally generalizes from the full conformal boundary to the boundary of the entanglement wedge.} This therefore enforces the $A'_1$ subsystem to live on the portion of the RT surface on the $A_1$ ``side'' of the $E_W$ surface, and similarly for the $A'_2$ and $A_2$ subsystems.

Let us now consider fixing an entanglement wedge $\rho_A$, and let us consider geometric partitionings of $A$ into $A_1$'s and $A_2$'s. Each chosen partitioning will give an $E_W$ surface, and assumption of the $E_P=E_W$ and surface-state conjectures will immediately give minimal Hilbert space dimension subsystems $A'_1$ and $A'_2$ associated with the geometric choice of partitioning. This gives an (admittedly restricted) version of the definition given in the previous section, as it suggests that the state defined on the entanglement wedge boundary given by the surface-state correspondence is, indeed the state that simultaneously calculates the entanglement of purification for all (geometric) partitionings of $A$. \footnote{It is noting, however, that the tension between one-shot and asymptotic entanglement distillation is eased in the holographic setting, as shown in the work of \cite{czech2015information}.} While this does not provide a statement regarding non-geometric partitionings of $A$, it is striking that $E_P=E_W$ naturally gives you a simultaneous extremization of geometric partitionings. This motivates a further definition of the holographic minimall entanglement purification of $\rho_A$.

\begin{defn}Given a density matrix of a geometric subregion of the boundary $\rho_A$, the holographic minimal entanglement purification is the state $\ket{\Psi}_{AB}$ that calculates the entanglements of purification associated with all geometric bipartitions of $\rho_A$. Furthermore, it is given by the state defined on the boundary of the entanglement wedge of $\rho_A$ in the surface-state correspondence.\end{defn}

Note that the condition on the Hilbert space dimension is guaranteed by the trivial paritioning of $\rho_A$ to an $A_1$ system and an empty $A_2$ system. Because holographic analysis in this case is limited to the cases of geometric partitionings of geometric boundary subregions $\rho_A$, it is not clear that the holographic minimal entanglement purification is also the minimal entanglement purification. If the $E_P=E_W$ conjecture is correct, however, it at a minimum has the right Hilbert space dimension and reproduces a large number of entanglements of purification that the minimal entanglement purification would produce. This motivates a proposition that these objects are, indeed, the same. 

\begin{prop}The holographic minimal entanglement purification is the minimal entanglement purification of a boundary geometric density matrix $\rho_A$.\end{prop}

Consider the partition of $A$ into any partitioning, $A_1$ and $A_2$. Again without loss of generality, let us consider only $A_2$. Let us repeat the distillation process, which together with the surface-state correspondence, this time pushing to surfaces with area corresponding to the entanglements from $A_1$ and $A_2$ from $A$, allows us to construct a pure state defined on the union of the RT surfaces of $A_1$, $A_2$, and $A$, where the state from $A$ was pushed to $A_1$ and $A_2$'s RT surfaces, and the state from the complement of $A$ was pushed to the $A$ RT surface. Note that the surface-state correspondence is general enough here that $A_1$ and $A_2$ need not be geometric; as long as their union is geometric, all of their entanglement data can be pushed onto some bulk subsurface with area equal to their respective entanglement entropies. As in the previous section, discussion of the $\bar{A}$ contribution to the entanglement of purification is fixed and irreducible, so we can focus only on the state defined on this new surface. The state defined on this new surface is geometric, so any entanglement of purification of any subset of it with any other subset will be given by the cross-section of the generalized ``entanglement wedge'' bounded by this surface. Because the claims that have been made now have been general for any boundary geometric density matrix $\rho_A$, it proves the proposition above.

Alternatively, one can simply take the existence and uniqueness of the entanglement wedge as proof of the existence and uniqueness of the holographic minimal entanglement purification.  As the constraints for minimal entanglement purifications are a superset of those for holographic minimal entanglement purifications, and the minimal entanglement purification is proven to exist, this sandwiches the holographic minimal entanglement purification to be the same as the minimal entanglement of purification.

The state in the surface-state correspondence used here is not a geometric state on the boundary; the $B$ portions of the state are quite nonlocally distributed in the complement of $\rho_A$. It would be nicer if there was some geometry for which $\ket{\Psi}_{AB}$ is naturally the state of a boundary geometric region. We find that there may be a way to do this in cases where $\rho_A$ is a subregion of pure locally $AdS_3$ spacetimes \cite{orbident}. In such cases, one can perform identifications to multiboundary wormhole geomerties for which the entanglement of purification surface becomes the throat calculating the entanglement entropies of some union of a subset of the complete boundaries defining the multiboundary wormhole. In this case, as the identification only maps the entanglement wedge and its boundary to the new wormhole geometry, the state of the new wormhole geometry should be $\ket{\Psi}_{AB}$. Moreover, as the identification does not change which ``side'' of the entanglement wedge cross section/throat the regions are on, and furthermore there are no more RT surfaces to localize to in the bulk, the $A_1A'_1$ and $A_2A'_2$ subsystems can only be unions of a subset of complete CFT boundaries respectively on either side of the throat. In this restricted set of cases, the geometrization implied by this conjecture is strengthened even further.

\section{A Comment on Hilbert Space Dimensionality}
Minimal entanglement purifications are best defined in finite dimensional Hilbert spaces, where differing infinities need not be extremized over. This is notably not the case, generically, in holographic CFT's, excluding the cases where the boundary subregions are entire CFT boundaries, where the RT surface area would be finite. In generic infinite dimensional cases, we are forced to rely heavily on the $E_P=E_W$ and surface-state conjectures, and on the fact that generically the RT surface area is much smaller than the total extensive volume of the boundary region it subtends. The role of the former is to give an independent holographic motivation for the minimality, at least in terms of total Hilbert space dimension of the purification, as it saturates the bound that the dimensionality of the purification equals the exponential of the entanglement entropy.

\section{Complexity=Action}
The holographic minimal entanglement purification conjecture could have interesting consequences for the complexity=action proposal \cite{Brown:2015lvg}. Briefly, this proposal states that the state complexity away from a reference state (normally taken to be the thermofield double state) of a boundary state is given by the action of the Wheeler-de Witt patch bounded by this boundary state. Furthermore, it has been proposed within this framework that increasing complexity corresponds to the existence of a firewall \cite{Brown:2017jil}.

The unconstrained version of this previous statement is clearly false; consider a pure state black hole of a given mass $M$ which we allow to evaporate over time. On general holographic grounds \cite{PhysRevD.7.2333, bousso1999covariant}, such a black hole would have a finite number of qubits associate with it, and thus a fixed maximum state complexity. If we augment this system with another much larger quantum system, and we increase the state complexity of this larger system, in general the overall state complexity will increase independent of what the black hole is doing. \footnote{We thank Raphael Bousso for first pointing this out to the author.} Thus, we would be led to the conclusion that a formally irrelevant (indeed, tensor-product) unitary to the black hole subsystem can affect whether or not there is a firewall present, which seems unlikely.

The minimal entanglement purification may be the correct restriction of state complexity to make this statement correct again. Let the black hole at a given time be the density matrix $\rho_A$. From this, one can consider the minimal entanglement purification for the black hole $\ket{\Psi}_{AB}$. If the holographic minimal entanglement purification conjecture is correct, then we can consider the entanglement wedge associated with the black hole. Here, we embed the black hole into an AdS-space to allow access to the holographic analysis. The entanglement wedge corresponding to the data from the CFT boundary of an AdS-Schwarzschild black hole is essentially ``half'' of a wormhole geometry, cut off at the throat. If one considers the purification of this state, the state from the surface-state correspondence would be then be state defined on the entire boundary union the state defined on the throat isometrically pushed to from the other conformal boundary. Now also apply the surface-state correspondence from the CFT side until one reaches a surface that has the same area as the black hole: this remaining geometry should contain all information about the black hole and its minimally entangled purification. Note that whether this pushes all the way to the horizon is controlled by the Page time, or e.g. whether the black hole entanglement entropy is yet equal to its area, something which is not true for sufficiently young black holes formed from the collapse of a pure state. The speculative claim here would be that only complexifying actions that would affect the state complexity of the minimal entanglement purification for the black hole can control whether or not a firewall exists, as all other complexifying actions can be taken to be formally irrelevant.

\begin{center} 
 {\bf Acknowledgments}
 \end{center}
 \noindent 
I thank Raphael Bousso, Grant Remmen, Arvin Shabazi Moghaddam, Aidan Chatwin-Davies and Illan F. Halpern for useful discussions.
N.B. is supported by the National Science Foundation under grant number 82248-13067-44-PHPXH and by the Department of Energy under grant number DE-SC0019380.

\bibliographystyle{utphys-modified}
\bibliography{Allrefs}

\end{document}